# Excitation caging in a vertex-frustrated quasiperiodic Einstein artificial spin ice

Tianyue Wang[1,2,*], Flavien Museur[1,2,*], Gavin M. Macauley[1,2,*,†], Jeanne Colbois[3], Luca Berchialla[1,2], Felix Flicker[4], Peter M. Derlet[1,5], and Laura J. Heyderman[1,2]

1. Laboratory for Mesoscopic Systems, Department of Materials, ETH Zurich, 8093 Zurich, Switzerland
2. PSI Center for Neutron and Muon Sciences, 5232 Villigen PSI, Switzerland
3. Institut Néel, CNRS UPR2940, 38042 Grenoble
4. School of Physics, University of Bristol, Bristol, BS8 1TL, UK
5. PSI Center for Scientific Computing, Theory and Data, 5232 Villigen PSI, Switzerland

[*] : Corresponding authors
[†] : Present Address: Department of Physics, Princeton University, Princeton, New Jersey 08540, USA

**Abstract**

Naturally occurring bulk quasicrystals are rare, and magnetic instances are rarer still, with chemical constraints typically permitting the synthesis of approximants rather than true quasicrystalline magnets. Here, we present an artificial spin ice based on a recently discovered Einstein lattice, the Hat tiling, which is built from the first known shape—the hat—that tiles the plane only aperiodically. The Einstein artificial spin ice has long-range structural order with no translational symmetry, and low-connectivity vertices with well-defined local ground states and excitations. Together, these properties provide a model two-dimensional quasicrystalline magnet, with magnetic correlations that we probe with magnetic force microscopy and parallel-tempered Monte Carlo simulations. We identify a two-stage partial ordering process, driven by the competition between two possible positions for magnetic excitations. This competition is resolved in the ground-state manifold, where exactly one magnetic excitation is caged on each antihat, yet the manifold remains macroscopically degenerate. This yields an unusual type of medium-range order, where the underlying quasiperiodic long-range order is randomly modulated by a strictly constrained disorder. Our findings establish the Einstein artificial spin ice as a blueprint for understanding quasicrystalline magnetism, demonstrating how quasiperiodic monotile geometries can be exploited to engineer unconventional magnetic phases with no direct equivalent in periodic systems.

**Introduction**

The field of crystallography developed out of the study of ordered atomic structures—including magnetic ones—that appear in nature, and the classification of their specific symmetries[1,2]. For centuries, the only known instances of long-range order were periodic. This dramatically changed in the 1980s with the discovery of *quasicrystals*, which are aperiodic structures with long-range order[3,4]. However, their magnetic properties have been explored less extensively[5–8], owing to the mathematical complexity created by the lack of translation symmetry but also to the scarcity of quasicrystalline magnetic minerals.

Artificial spin ices (ASIs) provide a way to bridge this gap. These are two-dimensional arrays of lithographically-designed elongated nanomagnets, which are coupled to each other principally through magnetic dipolar interactions[9–11]. In an ASI, the size and elongated shape of each nanomagnet are usually chosen so that they are single domain with the magnetization, or macrospin, pointing in one of two directions parallel to the nanomagnet long axis, thus giving a faithful representation of Ising variables. ASIs were initially developed as mesoscopic analogues of model frustrated magnets such as the rare-earth pyrochlores[12,13], enabling real-space characterization and straightforward tuning of the geometry. This led to the direct observation of exotic collective behaviours, including emergent magnetic monopoles[14,15], phase transitions[16–18], and vertex frustration[19–21].

While the focus has mainly been on periodic ASIs, there have been a few studies of quasicrystalline systems, often derived from the prototypical Penrose tilings[21–29]. For example, in Ref. 28, the authors found long-range magnetic order in an ASI based on the Penrose P1 tiling with dipolar interactions. In contrast, an interesting separation of energy scales was discovered in the Penrose P3 tiling with nearest neighbour interactions[21,29], which led to different behaviours for different sets of spins: a quasi-one-dimensional set, dubbed the skeleton, becomes long-range ordered at a high effective temperature, while an ensemble of flippable spins at vertices remains disordered down to low temperature. Similar clustering behaviour has also been observed in the bulk *i*-Tb-Cd quasicrystal with neutron scattering[30], where high-resolution measurements indicated evidence of inter-cluster magnetic correlation. However, in all these examples, it is the large connectivity of some of the vertices, rather than simply the quasiperiodic nature of the lattice, that leads to the observable physics.

At the same time, a long-standing problem in mathematics has been to find a shape that can tile the plane only aperiodically. This so-called aperiodic monotile problem, or “ein Stein” problem from the German for “one stone”, was solved by Smith *et al*. in 2023[31]. The solution is the *hat* monotile, that, together with its mirror image, the *antihat*, can be assembled only into the quasiperiodic (i.e. still long-range ordered) Hat tiling (see Fig. 1a). Soon after its discovery, the statistical mechanics of an Ising model with spins located on the vertices of the Hat tiling was investigated with Monte Carlo simulations[32]. This work established a critical temperature and Kramers-Wannier duality relations on a quasiperiodic lattice; but this spin system was not frustrated, with the ground state consisting of all spins pointing uniformly out of the plane. Addressing the case with in-plane spins, by placing dipolar-coupled nanomagnets on the edges of the Hat tiling to form an Einstein ASI, is therefore a perfect opportunity to study a model frustrated magnetic quasicrystal at the mesoscale.

Several features of the Hat tiling make it particularly attractive. First, the Hat tiling is only composed of vertices of low connectivity (3 and 4). Second, it is intimately connected to a periodic lattice (see Fig. S1), which guarantees that it has a hexagonal ($C_6$) rotational symmetry not found in most other quasiperiodic systems[33]. Finally, the ground state of systems with dipolar interactions is typically characterized by a flux closing of spins that align head-to-tail around the shortest loops of nanomagnets[10,34]. Because all hats in the Einstein lattice are the same, there is no preferred loop of head-to-tail macrospins that can form first, ensuring a highly degenerate landscape of configurations. Together, these properties make the energy levels, excitations and possible symmetry breakings of such an ASI simpler to define than for other quasicrystal lattices, thus providing a way to probe the correlations of a model quasiperiodic geometry.

Here, we fabricate an Einstein ASI based on the Hat tiling and show how its vertex structure and quasiperiodic nature shape exotic magnetic correlations across multiple length scales. Using Monte Carlo simulations, we identify several unique phenomena that have not been observed before in periodic and other quasiperiodic systems. Specifically, we find magnetic excitations in the low energy manifold that cannot be completely eliminated: the lattice topology prevents the nanomagnets at all vertices and edges from simultaneously adopting their lowest-energy spin configurations. Furthermore, we discover a ‘caging effect’, whereby these excitations are confined in specific antihat tiles but free to move between two positions on the tiles. We adopted this term in analogy with Aharonov–Bohm cages, where destructive interference bounds the set of sites that a quantum particle can visit[35]. On heating out of the ground state, the excitations first gain the freedom to move off antihats onto neighbouring edges, before fully delocalizing at higher temperatures. We therefore realize a phase that has no analogue in either the periodic or the Penrose-tiling-based ASIs. Beyond this, our work establishes a foundation for the design of other quasiperiodic-monotile ASIs and provides the necessary tools for characterising such systems.

**Construction of an Einstein ASI**

In Fig. 1a, we show the specific Einstein tiling, based on the hat tile, on which we build our ASI. The individual hats have four different colours according to their local environment: the antihats are shown in dark blue, and hats that belong to a *triskelion* motif (a six-hat motif with three radial arms) are shown in grey. The hats immediately adjacent to an antihat that do not belong to a triskelion motif are shown in light blue and all other hats are shown in white.

Using electron beam lithography, we fabricated an Einstein ASI with nanomagnets located on the edges of the hat tiles, as shown in the scanning electron micrograph in Fig. 1b. Each hat features 20 identical stadium-shaped permalloy nanomagnets with lateral dimensions of 450 nm × 150 nm. At these dimensions, each nanomagnet behaves as a single magnetic domain, with its magnetization pointing in one of two directions parallel to its long axis. Consequently, each nanomagnet can be represented by a macrospin Ising variable. Choosing a uniform size for all nanomagnets ensures all nanomagnets freeze at approximately the same temperature[36].

As illustrated in Fig. 1c, the Einstein lattice has a hierarchical structure: supertiles of larger order can be assembled by combining lower order supertiles. An individual hat constitutes the zeroth order tile (Supertile 0, here an antihat). As demonstrated in the original discovery[31], the choice of supertiles is not unique; nevertheless, antihats are a structurally important feature of the Hat tiling. For example, they have been shown to be central in understanding the electronic properties of a tight-binding model defined on an Einstein lattice[37]. Therefore, our choice of larger supertiles is motivated by ensuring that antihats always retain the same local environment. The first-order supertile (Supertile 1, outlined in green) involves an antihat surrounded by three hats. Then, three Supertile 1 structures are combined to form a second-order supertile (Supertile 2, outlined in purple), which, in turn, can be combined to form a third-order supertile (Supertile 3, outlined in yellow). The largest supertile that we studied, Supertile 4, is shown in Fig. S2.

In our Einstein ASI, either one or two nanomagnets are patterned along each edge of a hat depending on its length. In contrast to periodic ASIs, maintaining the same gap between all first nearest-neighbour nanomagnets is not possible when all nanomagnets are constrained to be the same size. Instead, we optimized the placements of individual nanomagnets along the hat edges to reduce the number of distinct gaps between first nearest-neighbour nanomagnets to three—with gap sizes of 16 nm, 50 nm and 120 nm—thus minimizing the variation in the magnitude of the dipolar interaction across the array.

The Einstein ASI has three different types of vertices, defined as locations where three or more nanomagnets meet, which are shown in orange in Fig. 1b. These are:

1. Cross-shaped vertices, involving four nanomagnets, in which one pair of nanomagnets across the vertex is more closely spaced than the other pair.
2. Y-shaped vertices involving three nanomagnets. There are two variants, with slightly different lattice constants.
3. T-shaped vertices, also involving three nanomagnets.

The energetics of the Einstein ASI are determined by two features: the macrospin configuration adopted by the nanomagnets at each vertex, and the alignment of macrospins in the nanomagnets along edges between vertices. Unlike most periodic ASIs, where both ends of each nanomagnet typically terminate at a vertex, there can be up to six nanomagnets arranged along an edge in our Einstein geometry. The macrospins of these edge nanomagnets strongly prefer to align head-to-tail. We refer to head-to-head or tail-to-tail arrangements of pairs of macrospins within the edges as edge excitations. We begin with characterising the magnetic configuration in experimental samples of increasing sizes, before providing a detailed description of the thermodynamics of this system.

**Increasing complexity of magnetic configurations in Einstein ASI supertiles**

As our Einstein ASI does not have a unit cell *per se*, we build our understanding of the static magnetic configurations in supertiles of increasing order. To allow each nanomagnet array to relax to a low-energy configuration, we apply a thermal annealing protocol in which the system is heated above the Curie temperature of the permalloy and then cooled back to room temperature. The resulting magnetic configuration is subsequently characterized using magnetic force microscopy (MFM, see Methods).

In Fig. 2, MFM images of the magnetic configurations in the supertiles of increasing order are shown alongside spin maps. In Fig. 2a, the MFM image of a single hat features all macrospins aligned head-to-tail in a closed loop, consistent with the expected dipolar ground state. In the larger Supertile 1 (Fig. 2b), formed from four hat tiles, we observe that the nanomagnets along the edges are in their ground state with the macrospins aligned head-to-tail, except for the nanomagnets at the two T-shaped vertices around the central antihat (indicated with orange dots), which are in their first excited state. Example macrospin configurations for all three vertex types in their ground state (GS) and each excited state (labelled E1, E2, E3 in order of increasing energy) are shown in Fig. 2c. In Supertile 1, it is impossible to construct a spin configuration in which all vertices are in their ground state and all edge interactions are satisfied. It follows that at least one residual excitation must remain, either along an edge or at a vertex (see Fig. S3).

The macrospin configuration of Supertile 2 in Fig. 2d shows signs of a higher-energy magnetic state. Six edge excitations, indicated with yellow crosses, can be observed, and are mostly located around the supertile boundary. In the bulk of the supertile, many excited vertices are present (shown with orange dots), including excited T-shaped vertices. The larger nanomagnet arrays that we studied experimentally (Supertile 3 and 4, see Fig. S2) also host both types of excitations.

To assess how close the experimental magnetic configurations are to the ground state, we perform Monte Carlo simulations to determine the ground states of the larger supertiles, assuming a point-dipolar model truncated to the nearest-neighbour level (see Methods). Similarly to Supertile 1, the simulated ground states retain a small number of excitations. For instance, the ground state of Supertile 2 hosts three excited T-shaped vertices, each located on one of the antihats (see Fig. 2e). Flipping macrospins to bring one of these T-shaped vertices into its ground state generates either an edge excitation or another excited vertex. The topology of the Einstein ASI thus prevents vertices and edges from all existing in their ground state simultaneously. We therefore expect that a certain number of local excitations is always present in the ground state, indicating that our system is frustrated. However, experimentally, we observe substantially more edge and vertex excitations in Supertile 2 and larger arrays than are predicted by the Monte Carlo simulations for their ground-state configurations.

Despite not reaching the ground state, the supertiles are far from completely disordered: the experimental magnetic structure factor (MSF, in the bottom left sector of Fig. 2f) shows sign of strong correlations. While there are no Bragg peaks, there are highly anisotropic diffuse patterns which reproduce the $C_6$ point symmetry of the Einstein ASI. Furthermore, the MSF is qualitatively very similar to that obtained for the Monte Carlo simulated ground state (upper left sector), with the powder average showing a good quantitative match (see Supplementary Fig. S4), corroborating that the truncation to a nearest-neighbour dipolar coupling captures the essential physics. A diffuse but structured MSF of this kind indicates that, while there is no long-range order, the experimental samples are strongly correlated in the short range.

**Two-stage thermodynamic crossover toward partial order**

To better understand how these correlations arise, we now turn to temperature-dependent Monte Carlo simulations of Supertile 4, using collective edge flips and parallel tempering (see Methods). The Monte Carlo simulations reveal two distinct crossovers in the specific heat capacity (Fig. 3a, indicated by red arrows): a broad maximum near $T \approx 30\ D/k_B$ and a smaller feature near $T \approx 3\ D/k_B$, where $k_B$ is the Boltzmann constant and $D$ is the dipolar energy constant (see Methods). We tracked the temperature dependence of the populations of the macrospin configurations associated with the different vertices (T-, Y- and cross-shaped), as well as the number of edge excitations. Y- and cross-shaped vertices go from a random distribution of energy levels at high temperature to having only the lowest energy level occupied at low temperature (see Fig. S5a, b for details). The higher temperature crossover is then straightforward to interpret: it marks the temperature at which most vertices—all cross shaped, all Y-shaped and most T-shaped vertices—simultaneously settle into their local ground-state configuration, which we refer to as the "local ice-rule".

The behaviour of the T-vertex populations (see Fig. 3b) and edge excitations (in Fig. 3c) is quite complex and must be interpreted together. At high temperature, the populations for the T-shaped vertex states also follow what is expected for a random spin configuration, with populations of 25%, 50% and 25% for the ground, first excited and second excited state, respectively. In addition, ~80% of edges (4467 out of 5644) host one or more excitation. As the system cools down toward $T \approx 10\ D/k_B$, the population of T-shaped vertices in their ground states rises to more than 90%, consistent with their local ice rule being satisfied almost everywhere, while the edge-excitation fraction drops below 2%. However, the trend inverts below $T \approx 10\ D/k_B$ as can be seen in Fig. 3b. The population of T-shaped vertices in their ground state decreases to about 80% while the first-excited T-shaped vertex population correspondingly grows and the number of edge excitations collapses to zero (Fig. 3c).

At first glance, this behaviour looks like population inversion, where a higher-energy state becomes more occupied as the temperature is lowered. However, here the population inversion happens at thermodynamic equilibrium, whereas it is normally observed in out-of-equilibrium systems where it can be described in terms of a "negative temperature"[38,39]. This apparent discrepancy is a consequence of considering vertex populations and edge excitations separately. However, they are intimately linked: as already mentioned in our discussion of Fig. 2, the suppression of an excitation on an edge creates an excited vertex and vice-versa. Since there must always be local excitations present, even in the ground state, our system thus exhibits vertex frustration[19–21]. To highlight this, we show in Fig. 3c that the total number of excited T-shaped vertices and edge excitations ($N_{\text{total excitations}}$, orange data points) is at a constant value of 81 in Supertile 4 below $T \approx 10\ D/k_B$ and corresponds to the residual number of excitations in the ground state that cannot be removed because of vertex frustration.

Commonly, vertex frustration becomes relevant around the scale of the nearest-neighbour interaction, in our case at $T \approx 10\ D/k_B$, corresponding to the first crossover in the specific heat capacity (Fig. 3a). However, when considering dipolar-coupled macrospins, it is slightly more favourable to have an excited T-shaped vertex than an edge excitation. This is why, around $T \approx 1\ D/k_B$, all edge excitations disappear while only excited T-shaped vertices remain.

This disappearance of edge excitations is responsible for the second crossover in the specific heat capacity. At intermediate temperatures between 1 and 10, excitations hop back and forth between T-shaped vertices and a neighbouring edge, while the total number of excitations remains constant. To conclude, at the lowest temperatures, all cross and Y-shaped vertices are in their ground states while no edge excitations remain. Hence the ground state manifold is characterized by the presence of excited T-shaped vertices that are only on the antihats.

We now compare the experimental vertex populations with those from Monte Carlo simulations to assign them an effective temperature. Eight Supertile 4 Einstein ASIs were thermally annealed, from which the average populations of the vertex types were obtained (see Methods). The closest agreement in the populations is obtained for an effective temperature of $k_B T/D \approx 25$ (see square points and vertical dashed line in Fig. 3 and Fig. S5): here, most vertices satisfy their local ice rule, but the vertex frustration is not established yet. The MSF computed at this temperature (right sector of Fig. 2f, labelled MC $T_{eff}$) now shows excellent quantitative agreement with the experimental MSF (bottom left sector in Fig. 2f, and Fig. S4). Our analysis confirms that quasiperiodic geometries can experimentally enforce complex magnetic correlations in artificial spin systems, yielding spin configurations situated between long-range order and disorder.

**Excitation caging and correlated disorder in the ground state**

In many ASIs, the ground state is composed of only the lowest-energy vertex configurations, and the true nature of the ground state is given by their correlations. We have shown that the ground state in Einstein ASI must host excited T-vertices on every antihat, and we now set out to give a complete description of their correlations. These excited T-shaped vertices can be located at one of two possible positions on the antihat (T1 and T2 in Fig. S6a). Describing their location is an *excitation localization problem*—considering the T-shaped vertex excitations only while disregarding the individual spins. We explain the exact nature of the resulting ground state and clarify why no phase transition is observed.

The positions of the excited T-shaped vertices in our simulated ground state are addressed at two length scales. For the long-range correlations, we compute the structure factors from the arrangement of excited T-shaped vertices only in the Monte Carlo simulations (shown in yellow in Fig. 4a), while for the short-range correlations, we compute their pair distribution function (yellow curve in Fig. 4d, see Methods). We compare these data to an ad-hoc correlated disorder model: excited T-shaped vertices are placed randomly at one of the T1 or T2 locations, with the constraint of one and exactly one excitation per antihat (see Fig. 4b). This correlated disorder model turns out to be in very good quantitative agreement with our numerical data at all length scales (compare green and yellow data in Figs. 4a and 4d). This agreement is seen more clearly in the differences between each of the three models and the numerical data in Fig. S7, for both the powder-averaged structure factors and the pair distribution functions. In this model, the ground state of the Einstein ASI is therefore a form of medium-range order[40], in which quasicrystalline long-range order is randomly modulated but subject to strong constraints. As no symmetry is broken, we do not expect a phase transition. Our description of the ground state bears a strong resemblance to that of an electronic tight-binding model on the Hat tiling, when a perpendicular magnetic field is applied[37]. There, the authors observed that the magnetic flux is localized at the antihats only in the ground state, indicating that this phenomenon is a general property of the physics of Hat tilings.

Nevertheless, there are small differences between the correlated disorder model and the Monte Carlo data, which could be the sign that the excitations are not only localized through the correlated disorder described above, but are also weakly coupled between neighbouring antihats through a small effective interaction. The effects of such an interaction can be understood with two limiting cases involving two complementary models: in the first case (shown in blue), we assume that all excited T-shaped vertices occupy the same position on the antihats. As the Einstein lattice is long-range ordered, the structure factor displays Bragg peaks representing the maximally ordered structure that can exist in the low-temperature manifold. If such a phase appeared spontaneously through the effect of small interactions between excitations, we expect it would be accompanied by a phase transition because of its long-range order. For the second case (shown in purple), we model a fully disordered "gas" of excited T-shaped vertices, placed randomly at the T1 and T2 locations, so that each antihat hosts between zero and two excitations (see Fig. 4c). The total number of excitations is fixed to match the total number of excitations in the Monte Carlo simulation. As expected in a disordered phase, there are no Bragg peaks in the structure factor. However, the pair distribution function shows that, in this case, the excitations are allowed to be much closer together than in the other models, with strong peaks around $r = 2$ μm. While these two models do not agree quantitatively with our simulated data, they demonstrate the richness of excitation localization regimes that can exist on the Einstein ASI.

Having established that the excited T-shaped vertices are mostly uncorrelated between antihats, we consider the consequences of setting the exact location of an excited vertex on an antihat and keeping all other vertices and edges in their ground state. This, in turn, sets the spin configuration up to time-reversal symmetry over a larger cluster of 51 spins. This cluster encompasses the antihat and adjacent regions, and we therefore refer to it as an *antihat cluster*. There are two distinct antihat cluster types, Type 1 and the doubly-degenerate Type 2 (Fig. S6a) with the excitation at position T1 and T2, respectively. Once the spin configuration of the antihat clusters is set (shown by red spins in Fig. S6b), the spins around the cluster borders (black spins in Fig. S6b) retain some configurational freedom. Indeed, all the border spins occur on edges between two Y-shaped vertices (shown in different colours in Fig. S8), which means that the entire edge can flip while retaining a ground-state configuration of the Y-shaped vertices so that there is no change in energy. Consequently, they never freeze—which contrasts with the long-range-ordered skeleton reported for the Penrose ASI[21,29]—and the system retains a macroscopic degeneracy.

We have identified the antihat clusters as a key building block of the nearest-neighbour ground state. It is natural to wonder whether this ground state persists when the full dipolar interaction is considered. Running Monte Carlo simulations with a long-range dipolar coupling would be too computationally expensive, but we can nonetheless predict its effect by comparing the exact dipolar energies of each antihat cluster state, which we show in Fig. S6a. State 2b, with the excited T-shaped vertex in the T2 position, has the lowest energy. In a full-range dipolar model at equilibrium, we can expect all antihat clusters to end up in this state, with the correlations described by the long-range ordered model (in blue in Fig. 4). Here, as the degeneracy of low energy states is explicitly broken, this is expected to result in a third crossover and not a phase transition.

Finally, the decomposition of the lattice into almost independent antihat clusters allows for an analytical estimation of the residual entropy. In the correlated disorder model, each antihat cluster carries six equivalent ground-state configurations: three distinct antihat cluster states (see Fig. S6a), each having two time-reversed variants. Then, the remaining nanomagnets on the border between antihat clusters can be grouped into several kinds of *border clusters*, defined as being an edge or set of edges between Y-shaped vertices (see Fig. S8). By considering almost independent clusters, while excluding the possibility of excited Y-shaped vertices, we estimate the ground state entropy of Supertile 4 to be $S_0^{indep}/N \approx 0.072\, k_B$ per spin (see Supplementary materials for details). The numerical value in the same system, obtained from integrating the specific heat capacity in our Monte Carlo simulations, is $S_0^{MC}/N \approx 0.062\ k_B$ per spin. Plotting the numerical residual entropy as a function of system size (see Fig. S9) indicates that this value is close to the infinite lattice value. The good agreement between our estimate and this numerical value confirms that the residual entropy is mostly due to the various degenerate local clusters that appear in our correlated disorder model. However, the difference in these residual entropies is another indication that the exact ground state manifold of the Einstein ASI is more complex than our simplified model. To remove this discrepancy, additional constraints in the border clusters should be introduced so that the local ice rule is satisfied everywhere. This would lead to small correlations between the antihat clusters and therefore between the positions of the excited T-shaped vertices. However, establishing the precise nature of these correlations requires more advanced simulation techniques and theoretical considerations that go beyond the scope of the present work.

**Conclusion**

In conclusion, we have established the Einstein ASI as a model quasiperiodic magnet, in which the ordering and thermodynamics are qualitatively distinct from that observed previously in both periodic and quasiperiodic ASIs. Specifically, two successive partial ordering regimes govern the thermodynamics on cooling: a high-temperature regime in which local ice rules are progressively obeyed across all vertex types, followed by a low-temperature regime in which the residual competition between edge excitations and excited T-shaped vertices is resolved. Rather than producing conventional long-range order, this yields a degenerate ground-state manifold in which exactly one magnetic excitation is localized on each antihat, forming extended correlated clusters, while spins bordering antihats have no identifiable correlations beyond the local ice rule.

The ground-state correlations are well captured by a correlated disorder model in which a single excitation resides randomly on one of the two T-shaped vertices on each antihat. The strong similarity between this ground state and that of a tight-binding model under field[37] suggests that the localization of zero modes around antihats is a general feature of Hat tilings. Small differences in the structure factor, pair distribution function, and residual entropy computed for an independent cluster model indicate the presence of weak inter-cluster correlations that warrant further investigation.

A degenerate ground state manifold is naturally very susceptible to small adjustments to the Hamiltonian. Therefore, modifying the Einstein ASI to maximize, or suppress, this degeneracy is a natural continuation of our work. Here, the presence of a finite number of excitations is due to the topology of the lattice, while the caging phenomenon relies on the fact that excitations can only diffuse by creating an excited vertex or edge. Thus, it should be possible to open the cages by tuning the vertex energies[18,41,42], so that the excitations can hop away from the antihats and interact with each other. In this case, we expect that the fully disordered model would better represent the system, at least over a finite range of temperatures. Alternatively, one could continuously add nanomagnets to the Einstein ASI to form its ordered underlying lattice (the deltoidal trihexagonal lattice shown in Fig. S1). This would result in a transition to a conventional long-range order as the lattice recovers its translational symmetry. Finally, the hierarchical structure of the Einstein lattice makes it well suited to techniques beyond Monte Carlo approaches[43,44], which could provide a more detailed understanding of the ground state and an analytical expression for the residual entropy. The Einstein ASI is therefore an important platform to explore and comprehend magnetic ordering and excitations in quasicrystalline or amorphous systems.

**Author contributions**

T.W., F.M. and G.M.M. conceived the project, analysed the data, established the story, and wrote the manuscript. T.W. fabricated the samples, performed the experiments and prepared the figures. G.M.M. wrote the analysis code to interpret the MFM images. P.M.D. wrote and ran the parallel-tempered Monte Carlo simulations. P.M.D., F.F. and J.C. provided theoretical and modelling support. L.J.H. supervised the work. All authors reviewed and commented on the draft.

**Acknowledgements**

This work was funded by the Swiss National Science Foundation (Grant Numbers 200332 and 10005640). We thank the cleanroom operations teams of the Paul Scherrer Institute for their help and support.

## Methods

### Fabrication

Our samples consist of arrays of elongated permalloy nanomagnets arranged in five different Supertiles 0 to 4 where Supertile 0 is the single hat. The samples were prepared by spin-coating polymethyl methacrylate resist with a thickness of 70 nm on a 10 mm × 10 mm silicon substrate. The resist was then patterned using a Vistec EBPG 5000PlusES electron beam writer at 100 keV accelerating voltage. The area of resist exposed to the electron beam was removed in a developer consisting of a mixture of methyl isobutyl ketone and isopropanol (IPA) in the ratio 1:3 for 45 s before being cleaned with isopropanol for 30 s. To create the nanomagnets, a 9 nm-thick film of permalloy (Ni 80%, Fe 20%) was deposited by thermal evaporation at a base pressure of $2 \times 10^{-6}$ mbar. The permalloy layer was deposited at a rate of 1.6 $\text{Å}s^{-1}$ and capped to prevent oxidation with a 2 nm-thick layer of aluminium deposited at a rate of 0.4 $\text{Å}s^{-1}$. The thickness of permalloy was chosen to give a relatively low energy barrier for magnetization switching while being thick enough to give sufficient signal for MFM imaging. The remaining resist with unwanted magnetic material was removed in hot acetone at 40 ℃ and then hot dimethyl sulfoxide at 50 ℃, both in an ultrasonic bath, before the sample was ultrasonicated for 3 minutes in IPA as the final cleaning step. The resulting lateral dimensions of the individual nanomagnets were 450 nm in length by 150 nm in width. The lattice constant is 500 nm, which we defined as the length of the shortest edge of an Einstein hat, and was chosen to maximize the coupling strength by minimizing the gaps between magnets while ensuring that the nanomagnets do not touch.

### Annealing

The thermal annealing protocol was conducted in an AJA Sputtering tool with a base pressure of $3.6 \times 10^{-8}$ mbar. The sample was attached to the sample holder using silver paste to give good thermal contact. The sample was heated from room temperature to 850 ℃ at a rate of 0.2 $℃s^{-1}$, above the Curie temperature of permalloy of about 600 ℃. The sample was kept at this temperature for 15 minutes before it was cooled to room temperature at a rate of 0.1 $℃s^{-1}$. We used the slowest cooling rate and maximum temperature possible in the sputtering tool.

**Magnetic Force Microscopy**

The sample was imaged using a Bruker Dimension 3100 magnetic force microscope, employing low-moment tips MESP-LM-V2 to prevent switching of the magnetic moments. The imaging parameters were identical for imaging the supertiles of different orders: a scan size of $15\ \mu\mathrm{m} \times 15\ \mu\mathrm{m}$, 512 scan lines, and a scan rate of 1 Hz. The magnetic contrast was obtained with the lift mode and a 40 nm lift height. For supertiles larger than the scan size, the full patterns were reconstructed by stitching multiple adjacent scans with consistent scanning parameters and a region of overlap to ensure accurate alignment.

**Monte Carlo Simulations**

To explore the thermodynamics and the low-energy states of the Einstein ASI, we conducted Monte Carlo simulations by treating each magnet as an Ising macrospin located at the centre of a magnet and pointing in one of two directions parallel to the long axis of the magnet. Magnets are coupled to each other through the magnetic dipolar interaction, described by the Hamiltonian

$$\mathcal{H} = D \sum_{\langle i,j \rangle} \frac{1}{{r_{ij}}^3} \left[ \boldsymbol{\mu}_i \cdot \boldsymbol{\mu}_j - 3(\boldsymbol{\mu}_i \cdot \hat{\mathbf{r}}_{ij})(\boldsymbol{\mu}_j \cdot \hat{\mathbf{r}}_{ij}) \right],$$

where $\hat{\mathbf{r}}_{ij}$ is the unit vector connecting the centres of macrospins $i$ and $j$, $r_{ij}$ is the distance between the two macrospins, and for spin $i$, $\boldsymbol{\mu}_i$ is the magnetic moment defined as $\boldsymbol{\mu}_i = \sigma_i \hat{\boldsymbol{e}}_i$ (where $\sigma_i = \pm 1$ and $\hat{\boldsymbol{e}}_i$ is the unit vector along the nanomagnet long axis). In our simulations, we assume the same distances between macrospins as in our experimental samples, meaning that all distances are expressed in $\mu\mathrm{m}$. The dipolar energy constant $D$ is defined as

$$D = \frac{\mu_0}{4\pi} \frac{m^2}{a^3}.$$

Here, the magnitude of the macrospin moment is defined as $m = M_S V$, where $M_S$ is the saturation magnetization, $V$ is the volume of the magnets, and $a$ is the lattice parameter given by the length of one side of the hat that is defined to have unit length. To limit the runtime of simulations, we truncate the dipolar coupling to nearest neighbours, an approach widely used for ASIs[45,46]. In our case, given the sparse geometry of the Einstein tiling and the rapid $1/r^3$ decay of dipolar interactions, this accurately captures the essential physics.

High-temperature ($k_B T/D > 10$) simulations were run with single spin flip dynamics and a Metropolis acceptance criterion. Below this temperature, the energy landscape starts to exhibit many local minima which prevent the convergence of the single spin flip algorithm in a reasonable time. Therefore, low-temperature($k_B T/D \leq 10$) simulations were run using a parallel tempering algorithm[47] to ensure that the system efficiently samples the phase space. In this method, an ensemble of replicas of the system at different temperatures is simulated together. Then, configurations are exchanged between replicas at neighbouring temperatures according to the Metropolis criterion. Its effectiveness relies on the fact that configurations at high temperatures become available to the replicas at low temperatures, and vice versa. After the exchange of configurations between replicas, the dynamics within each replica proceeds according to a combination of single spin and collective spin flips. In the latter, an edge is randomly selected and all spins belonging to the selected edge are flipped simultaneously. One Monte Carlo sweep was defined as $N$ single spin flip trials and $0.1N$ collective flip trials, where $N$ is the number of spins. To obtain converged results over the broad temperature range $k_B T/D \in [0.01, 10]$, the parallel tempering method was employed on a set of 60 logarithmically spaced temperatures spanning this range.

**Magnetic structure factor (MSF) and its powder average**

For a system of $N$ Ising macrospins, each with a magnetic moment $\boldsymbol{\mu}_i$ located at position $\boldsymbol{r}_i$, the MSF is defined as:

$$S_m(\boldsymbol{q}) = \frac{1}{N}\left|\sum_{i=1}^{N}[\boldsymbol{\mu}_i - (\boldsymbol{\mu}_i \cdot \hat{\boldsymbol{q}})\hat{\boldsymbol{q}}]e^{i\,2\pi\,\boldsymbol{q}\cdot\boldsymbol{r}_i}\right|^2,$$

where $\hat{\boldsymbol{q}} = \frac{\boldsymbol{q}}{|\boldsymbol{q}|}$ is a unit scattering vector. To reflect what is observed in neutron scattering experiments[48], only the projection of the magnetization perpendicular to the scattering vector contributes to the scattering amplitude. The MSF was evaluated in the $(q_x, q_y)$-plane on a 2000 × 2000 grid spanning $q_x, q_y \in [-1.5, 1.5]\ \mu m^{-1}$. To suppress sampling artefacts associated with the finite lattice and produce smooth intensity maps, the resulting $S_m(\boldsymbol{q})$ was smoothed with a Gaussian kernel of width σ = 5 pixels.

To quantitatively compare the MSFs from experiments and Monte Carlo simulations, powder averages were computed from

$$S_m(q) = \frac{1}{2\pi}\int_0^{2\pi} S(q\cos\phi\,, q\sin\phi)d\phi,$$

giving the radial average of a 2D MSF over all orientations on the ring $\boldsymbol{q} = (q\cos\phi\,, q\sin\phi)$.

**Excited T-shaped vertices positions: structure factor and pair distribution function**

We computed the structure factor of the positions of excited T-shaped vertices, treating each excited vertex as an identical point scatterer:

$$S(\boldsymbol{q}) = \frac{1}{N}\left|\sum_{j=1}^{N} e^{i\,2\pi\,\boldsymbol{q}\cdot\boldsymbol{r}_j}\right|^2,$$

where the sum now runs over the $N$ excited T-shaped vertex positions in each configuration. The same grid, smoothing, and normalization procedure as for the MSF was applied.

To compute the pair distribution function, all pairwise distances between excitations $r_{ab} = |\boldsymbol{r}_a - \boldsymbol{r}_b|$ were collected up to a cutoff $r_{max} = 10\ \mu m$, and the number of pairs of excited T-shaped vertices at each distance $n(r)$ counted. The pair distribution function

$$g(r) = \frac{n(r)}{N\,\rho\,2\pi\,r}$$

gives a measure of how the probability of finding excited T-shaped vertices at a certain distance $r$ differs from that of an ideal gas of the same density. If $g(r)$ is larger (smaller) than 1, the probability of finding excitations separated by a distance r is larger (smaller) than in an ideal gas. The average density of excitations is $\rho = N/A$, with $N$ the total number of T-shaped vertices, and $A$ the area of the system. In a disordered phase, the correlations decay as r increases, so $g(r)$ tends to 1.

Finally, the excitations are located on a discrete lattice, which means that the raw $g(r)$ are a combination of delta functions making comparison difficult. Therefore, a uniform Gaussian smoothing was applied to all pair distribution functions.

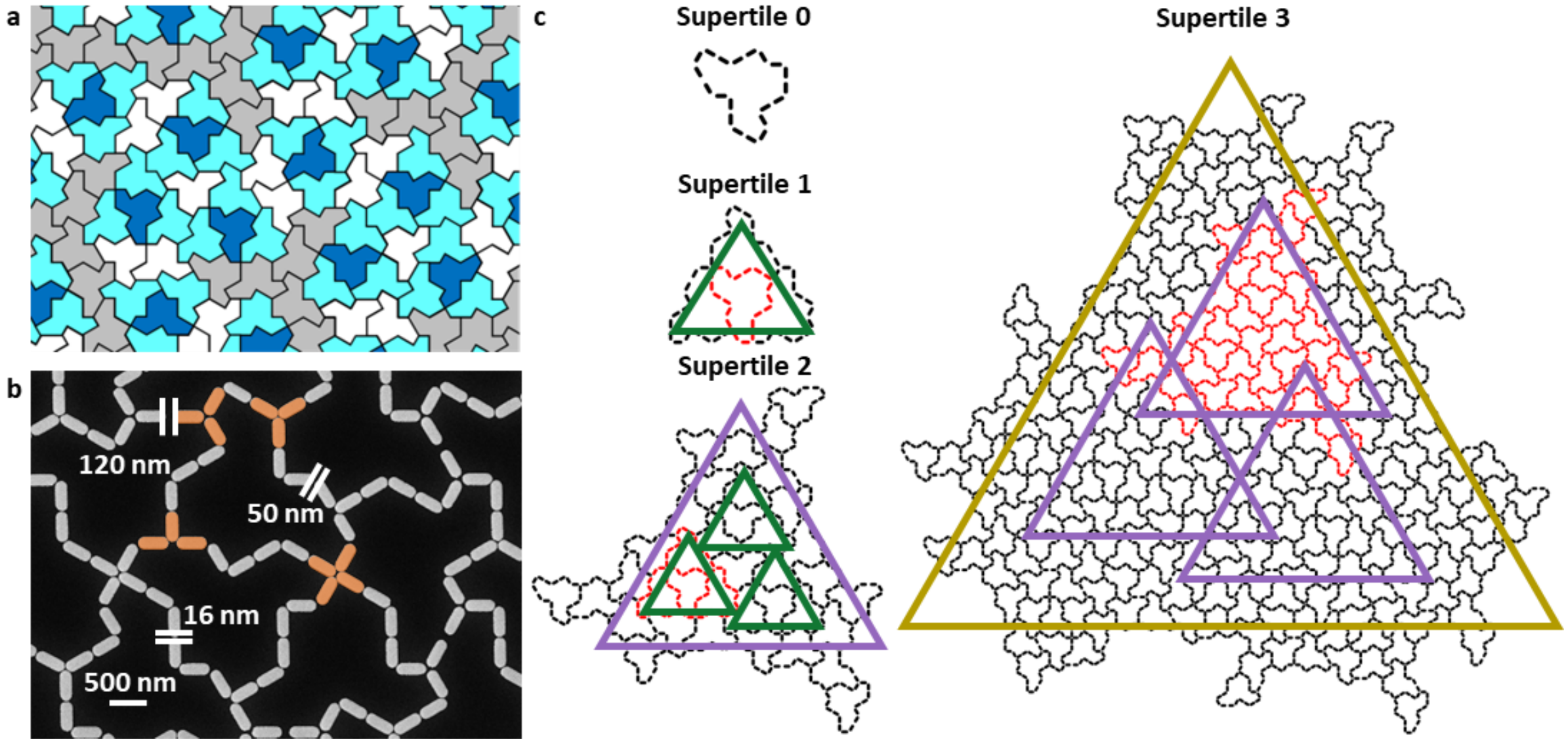

**Fig. 1 | Geometry, vertex definition and hierarchical construction of the Einstein ASI. a,** The quasiperiodic Hat tiling is composed of a single type of tile, called the hat. Hats are coloured according to the motif to which they belong: dark blue for antihats, which are the mirror images of the hat; grey for hats forming a three-legged triskelion; light blue for hats surrounding an antihat; and white for all remaining hats. **b**, Scanning electron micrograph of an Einstein ASI based on the Hat tiling. Nanomagnets, of length 450 nm and width 150 nm, are located along each edge of the Hat tiles, with one or two nanomagnets placed along each edge depending on the edge length. The positions of the nanomagnets have been carefully chosen to reduce the number of distinct gaps between adjacent nanomagnets to three: 16 nm, 50 nm, and 120 nm, as indicated. The three vertex types are indicated in orange: a four-nanomagnet squashed-cross vertex, a three-nanomagnet T-shaped vertex, and two variants of three-nanomagnet Y-shaped vertices, distinguished by their different gaps. **c**, The Einstein ASI is built from a hierarchical construction in which larger supertiles are built from smaller ones. In this framework, the Hat tile can be considered Supertile 0. The first four supertiles are shown, all of which are approximately triangular in shape. Each supertile forms the building block of the next order, as highlighted by the coloured triangles. Supertile 4 is shown in Fig. S2.

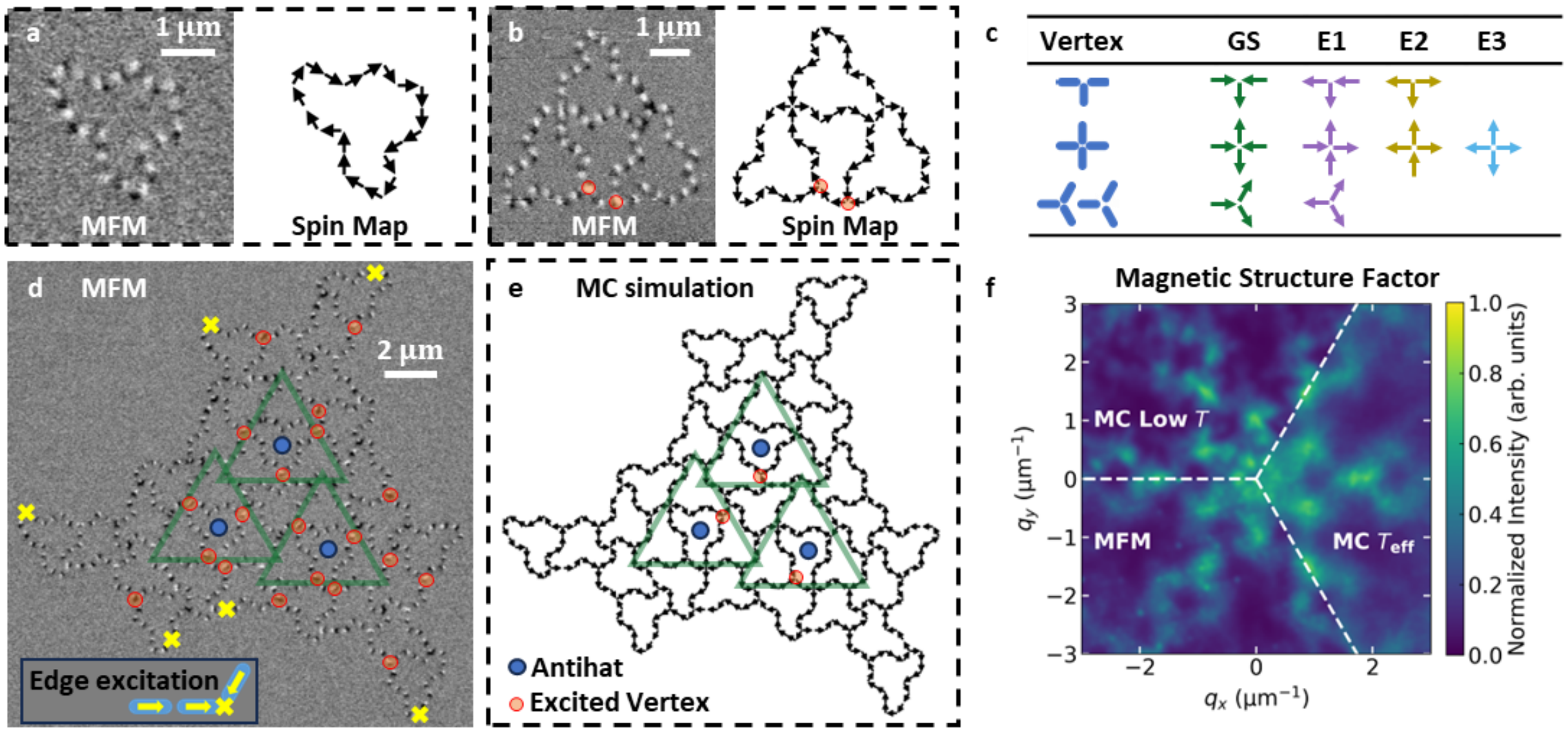

**Fig. 2 | Magnetic configurations of thermally annealed Einstein ASIs of increasing size. a**, MFM image and extracted spin map for a single hat. All of the macrospins are aligned head-to-tail, giving the expected ground-state configuration. **b**, MFM image and corresponding spin map for Supertile 1. **c**, Example vertex configurations for each energy level of the three vertex types: T-shaped, cross-shaped, and Y-shaped. **d**, MFM image of Supertile 2. Blue dots indicate the locations of the antihats, and green triangles indicate Supertile 1 building blocks. Edge excitations—defined as locations where two macrospins are arranged head-to-head or tail-to-tail—are identified with yellow crosses. A schematic (top left) illustrates a representative edge excitation as a yellow cross on a three-spin edge. **e**, Ground state of Supertile 2 from Monte Carlo (MC) simulations. In panels **b**, **d** and **e**, vertices in excited states are denoted by orange dots. **f**, Magnetic structure factor (MSF) of Einstein ASIs on Supertile 4. The three sectors compare the MSFs from the MFM experimental data (bottom left) with those obtained from Monte Carlo simulations in the ground state (top left) and at the effective temperature (right). The Monte Carlo MSFs were each averaged over 100 independent simulated configurations. The experimental MSF was averaged over the spin configuration of 8 thermally annealed arrays.

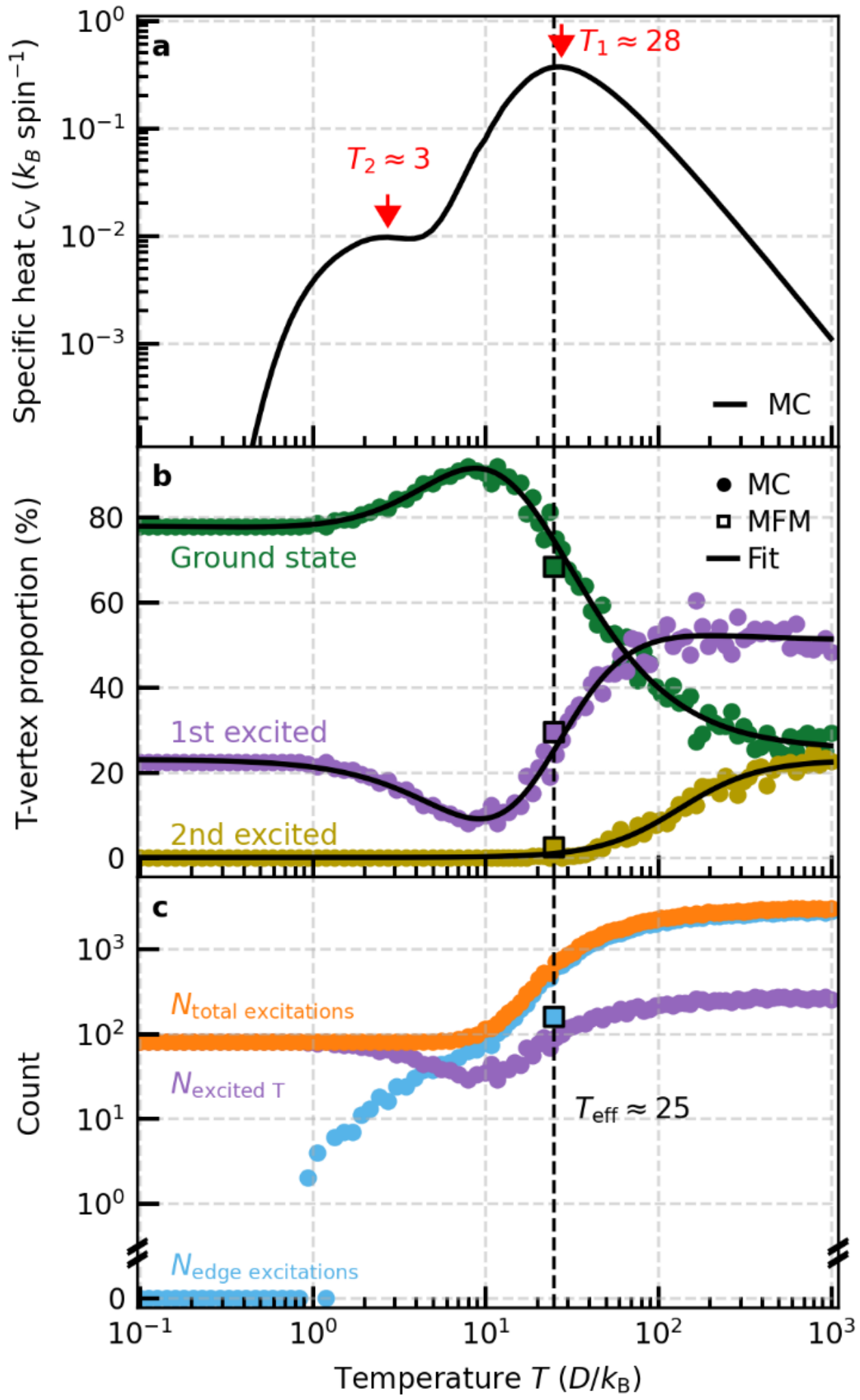

**Fig. 3 | Two-stage thermodynamic ordering process and vertex frustration. a**, Specific heat capacity computed from Monte Carlo (MC) simulations for Supertile 4, with 7633 nanomagnets. **b**, Populations of T-shaped vertices in the ground (green), first-excited (purple), and second-excited (yellow) states as a function of temperature. Dots correspond to the simulated populations, while squares denote experimental data extracted from MFM. Solid lines represent logistic fits to the simulation data. **c,** Number of edge excitations versus temperature (blue). The total number of low-temperature excitations, defined as the sum of the number of edge excitations and the number of T-shaped vertices in their first excited state (purple), is shown in orange: the constant non-zero value at low temperature signals vertex frustration. The dashed vertical line indicates the estimated effective temperature of the experimental data, where the observables were averaged from the spin configurations of 8 annealed Supertile 4 arrays.

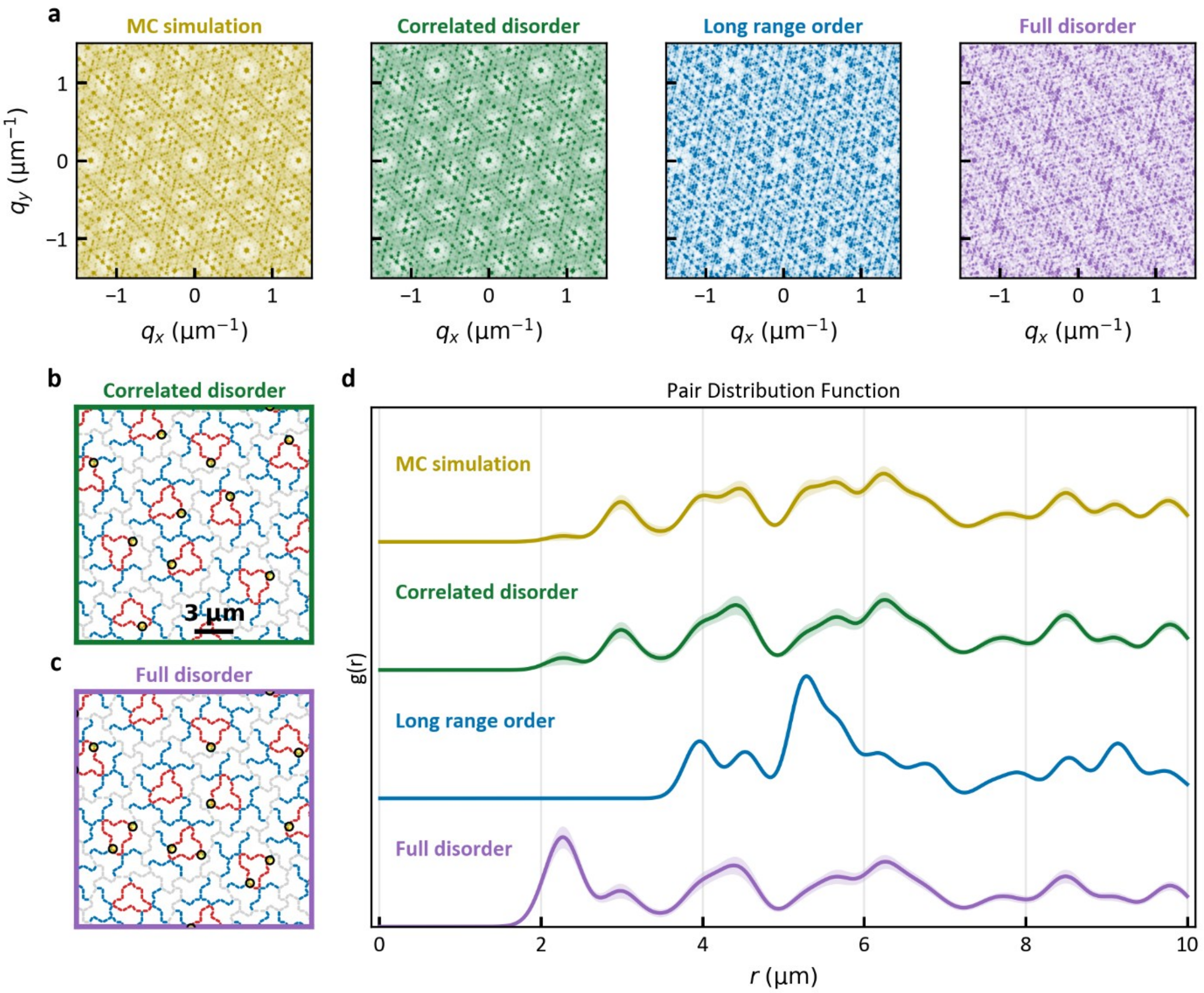

**Fig. 4 | Ground state correlations between the numerical data and several models at short- and long-length scales. a**, Structure factors of excited T-shaped vertices. The four graphs display the structure factor for: the simulated Monte Carlo ground-state manifold (yellow); a correlated disorder model (green); a maximally ordered configuration where every antihat hosts an excitation at the T1 position (blue), with locations of T1 and T2 given in Fig. S6a; and a fully disordered case (purple). **b-c**, Real-space schematics illustrating the two disordered models. Red and blue magnets represent the antihats and their associated clusters, respectively; all other magnets are shown in grey. Selected T-shaped vertex excitations are highlighted by yellow dots. **d**, Pair distribution function for the four models, calculated up to a radius of $r = 10\ \mu\text{m}$. The Monte Carlo, correlated disorder and full disorder structure factors and pair distribution function were each averaged over 100 independent configurations.

**Excitation caging in a vertex-frustrated quasiperiodic Einstein artificial spin ice:**

**Supplementary Information**

Tianyue Wang[1,2,*], Flavien Museur[1,2,*], Gavin M. Macauley[1,2,*], Jeanne Colbois[3], Luca Berchialla[1,2], Felix Flicker[4], Peter M. Derlet[1,5], and Laura J. Heyderman[1,2]

1. Laboratory for Mesoscopic Systems, Department of Materials, ETH Zurich, 8093 Zurich, Switzerland
2. PSI Center for Neutron and Muon Sciences, 5232 Villigen PSI, Switzerland
3. Institut Néel, CNRS UPR2940, 38042 Grenoble
4. School of Physics, University of Bristol, Bristol, BS8 1TL, UK
5. PSI Center for Scientific Computing, Theory and Data, 5232 Villigen PSI, Switzerland

[*] : Corresponding authors

[†] : Present Address: Department of Physics, Princeton University, Princeton, New Jersey 08540, USA

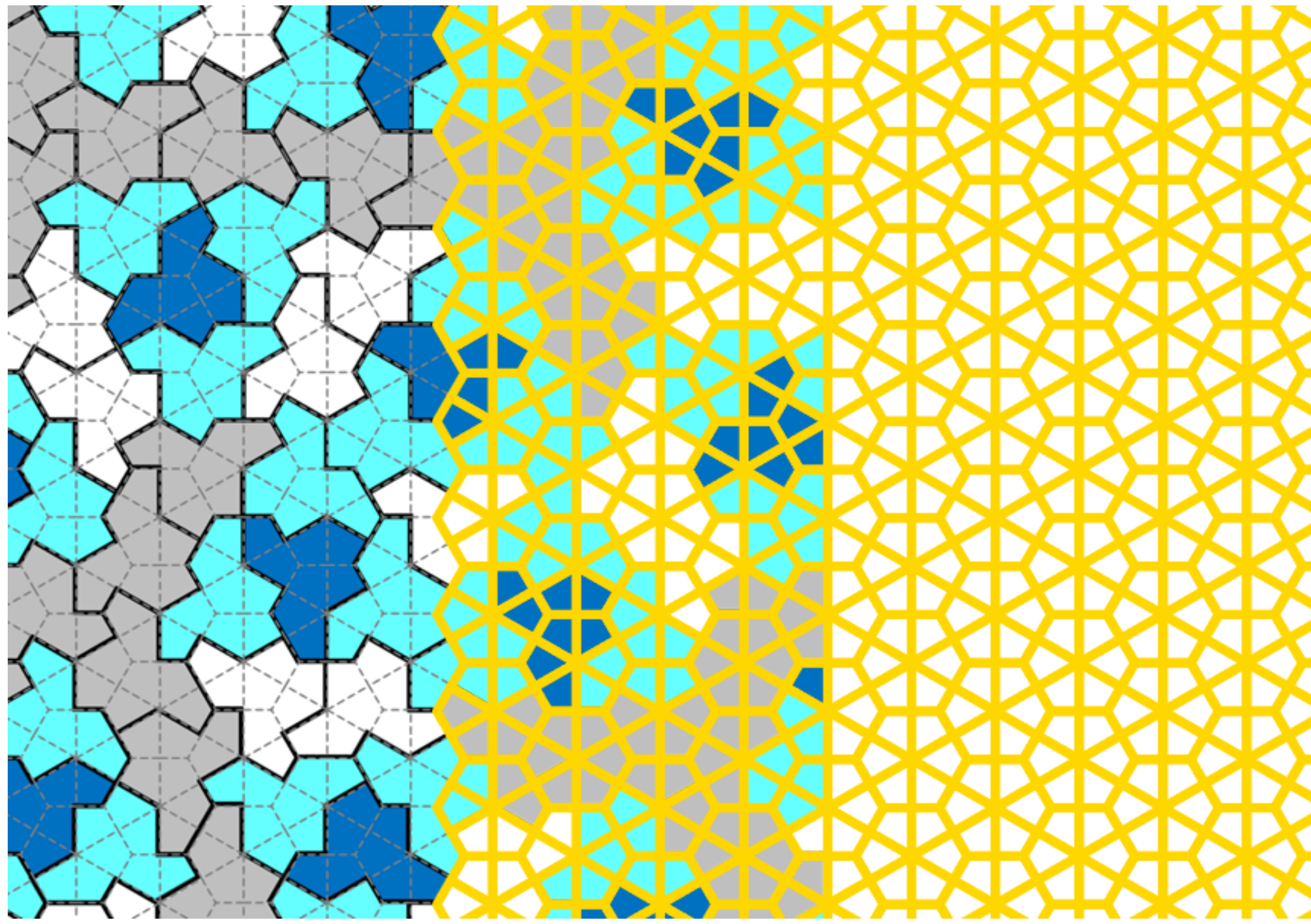

**Fig. S1 | The Hat tiling and its underlying periodic tiling.** The Hat tiling is delineated by the solid black lines on the left, with the underlying periodic *deltoidal trihexagonal tiling* indicated with dashed lines. This periodic tiling is further highlighted using yellow solid lines to the right, overlapping the Hat tiling in the middle of the figure.

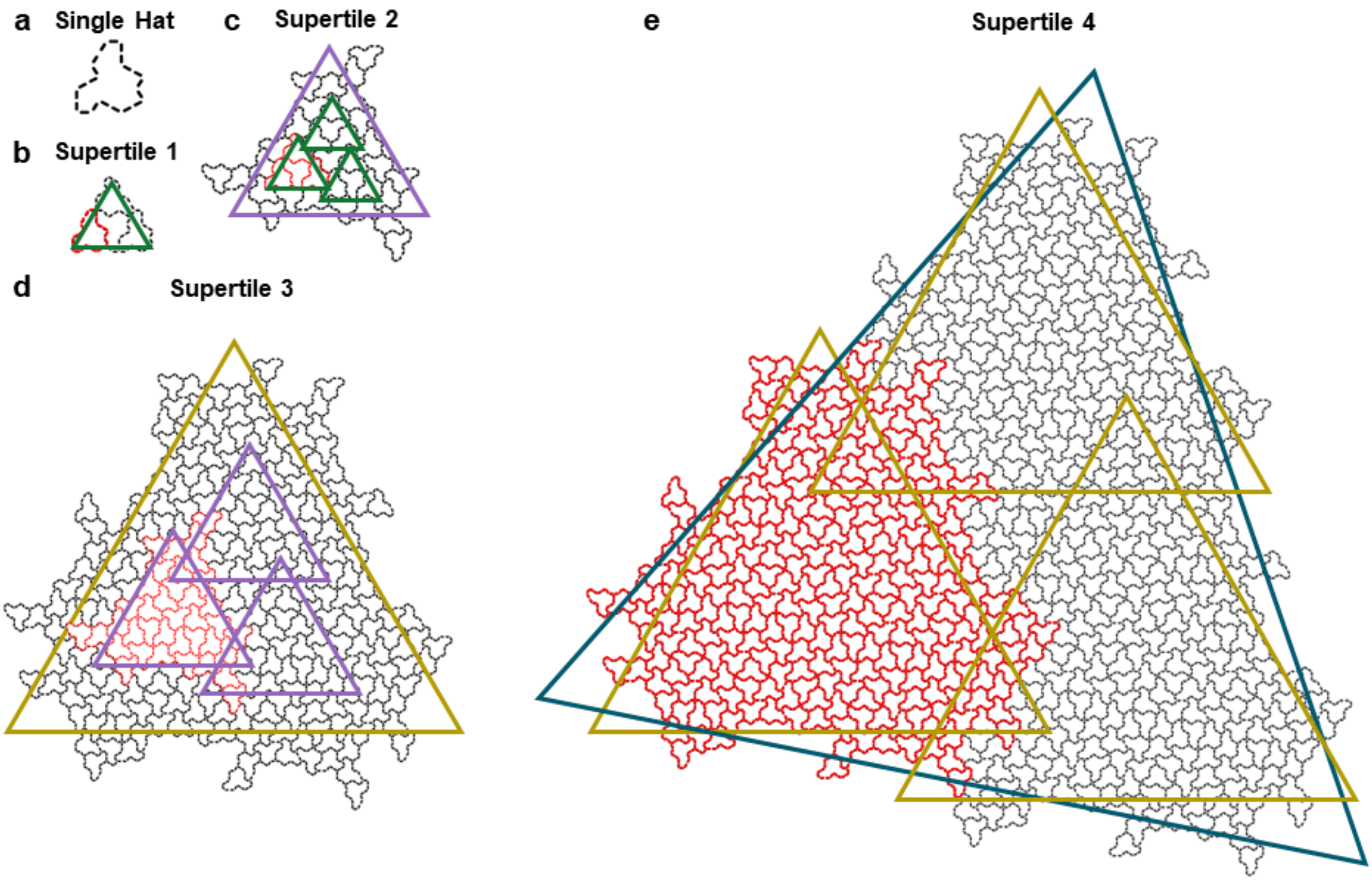

**Fig. S2 | Supertiles of the Einstein artificial spin ice.** The first five supertiles are shown, going from Supertile 0 (**a**), which is a single hat, to Supertile 4, which is an artificial spin ice with 763 hats (**e**). Each Supertile is triangular in shape and is highlighted in red in the next generation of supertile.

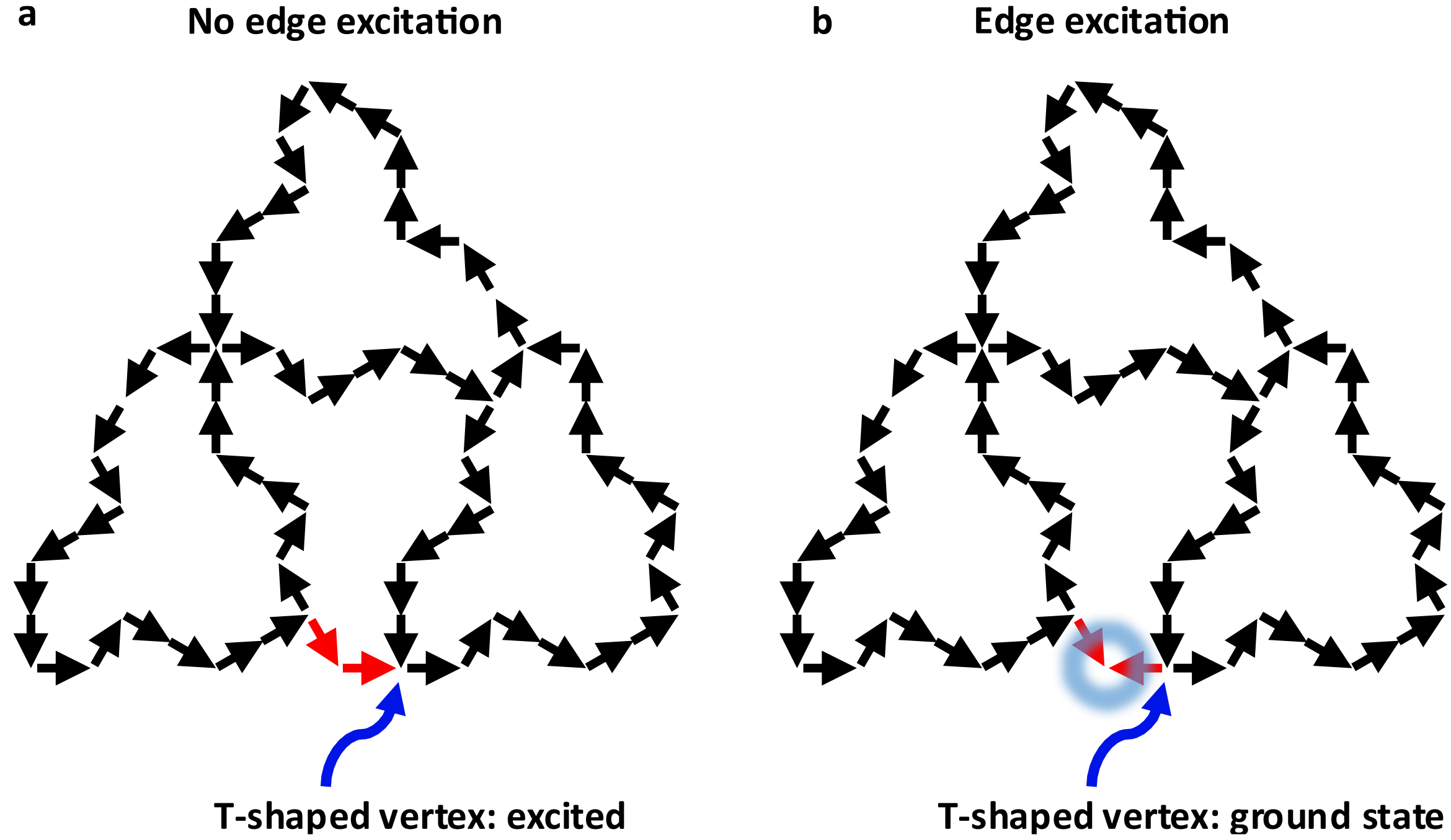

**Fig. S3 | Low temperature spin configurations in Supertile 1. a**, This configuration contains no edge excitation, and residual excitation is at the excited T-shaped vertex indicated by the blue arrow. **b**, This configuration has the edge excitation highlighted by the blue ring as the residual excitation, and the T-shaped vertex indicated by the blue arrow is now in its ground state. The array of macrospins can switch between these two configurations by flipping the horizontal red spin at the bottom.

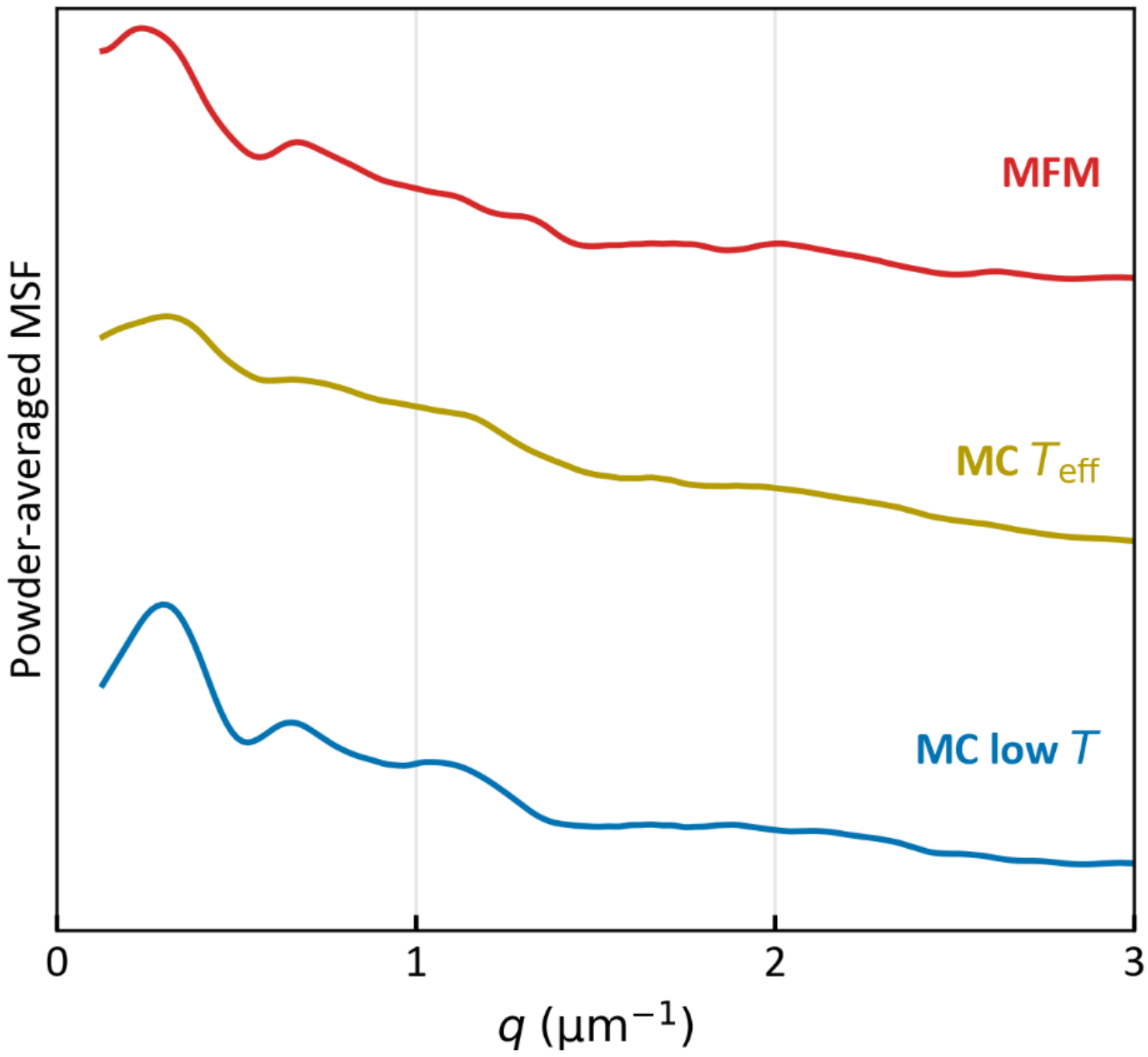

**Fig. S4 | Powder-averaged Magnetic Structure Factors.** These correspond to the three MSFs displayed in Fig. 2f and are determined from the experimental data obtained by MFM (in red), from the Monte Carlo simulation data at the effective temperature (in yellow), and from the Monte Carlo simulation data for the ground state (in blue). At low temperatures, we observe stronger correlations at small $q$ for the Monte Carlo simulation data for the ground state, which do not appear in the experimental MFM and in the simulated data at the effective temperature.

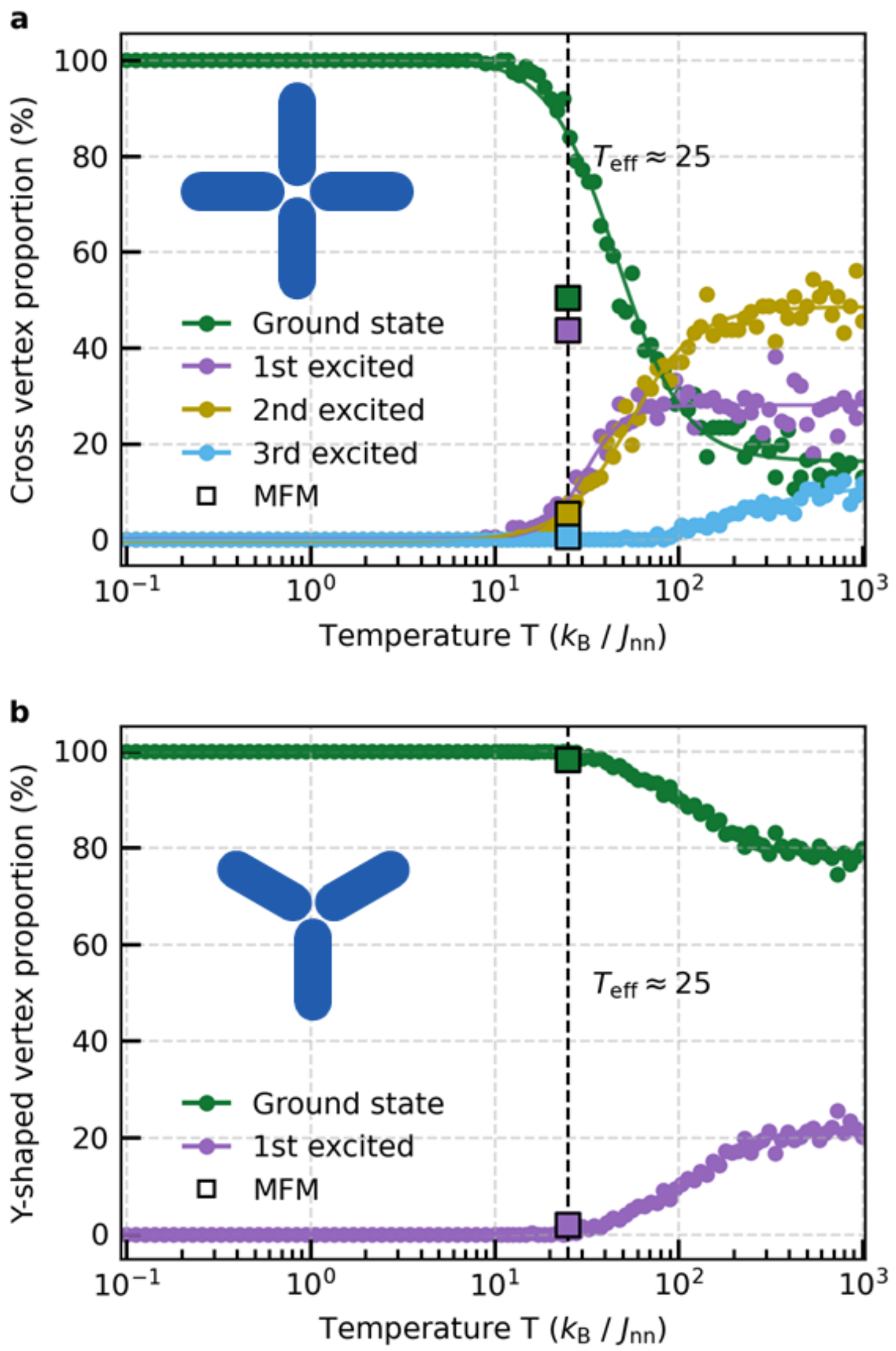

**Fig. S5 | Populations of vertices in different energy states as function of temperature obtained from Monte Carlo simulations.** Populations of **a**, cross-shaped vertices and **b**, Y-shaped vertices in the ground (green), first-excited (purple), second-excited state (yellow), and third-excited state (blue) as a function of temperature. The energy difference between the ground state and first excited state of the cross vertices is small, meaning that the populations can be easily influenced by a small bias field or another external perturbation. This explains the larger mismatch in effective temperature for this vertex type.

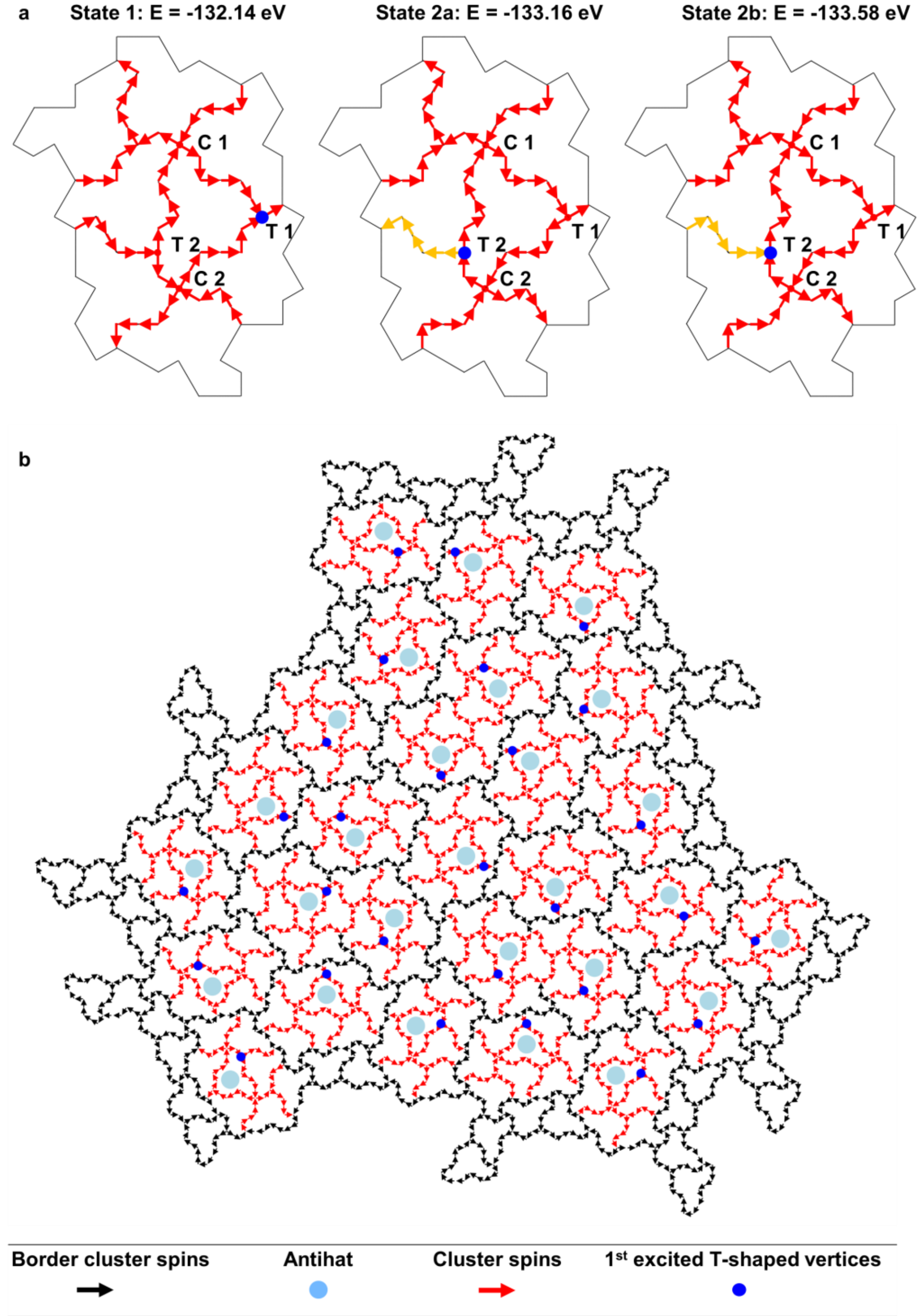

**Fig. S6 | Antihat cluster states and ground state manifold. a**, Antihat cluster (red and yellow arrows) in three different states. While all states have the same energies when computed with nearest-neighbour interactions, this is not the case when the longer-range part of the interaction is included. The energies of the different states, computed for the full range of dipolar interactions, are displayed and increase when going from State 1 to State 2a to State 2b. The excited T-shaped vertex is highlighted by a blue dot located at T1 or T2. State 2b is obtained by switching the yellow macrospins in State 2a, with the excitation remaining at the same location. **b**, Example ground state for Supertile 3. Antihat clusters of spins, which are indicated with red arrows, are separated by non-cluster spins in black. The T-shaped vertex excitations, which are caged by the antihats, are highlighted with small dark blue dots and the antihats are indicated with large light blue dots.

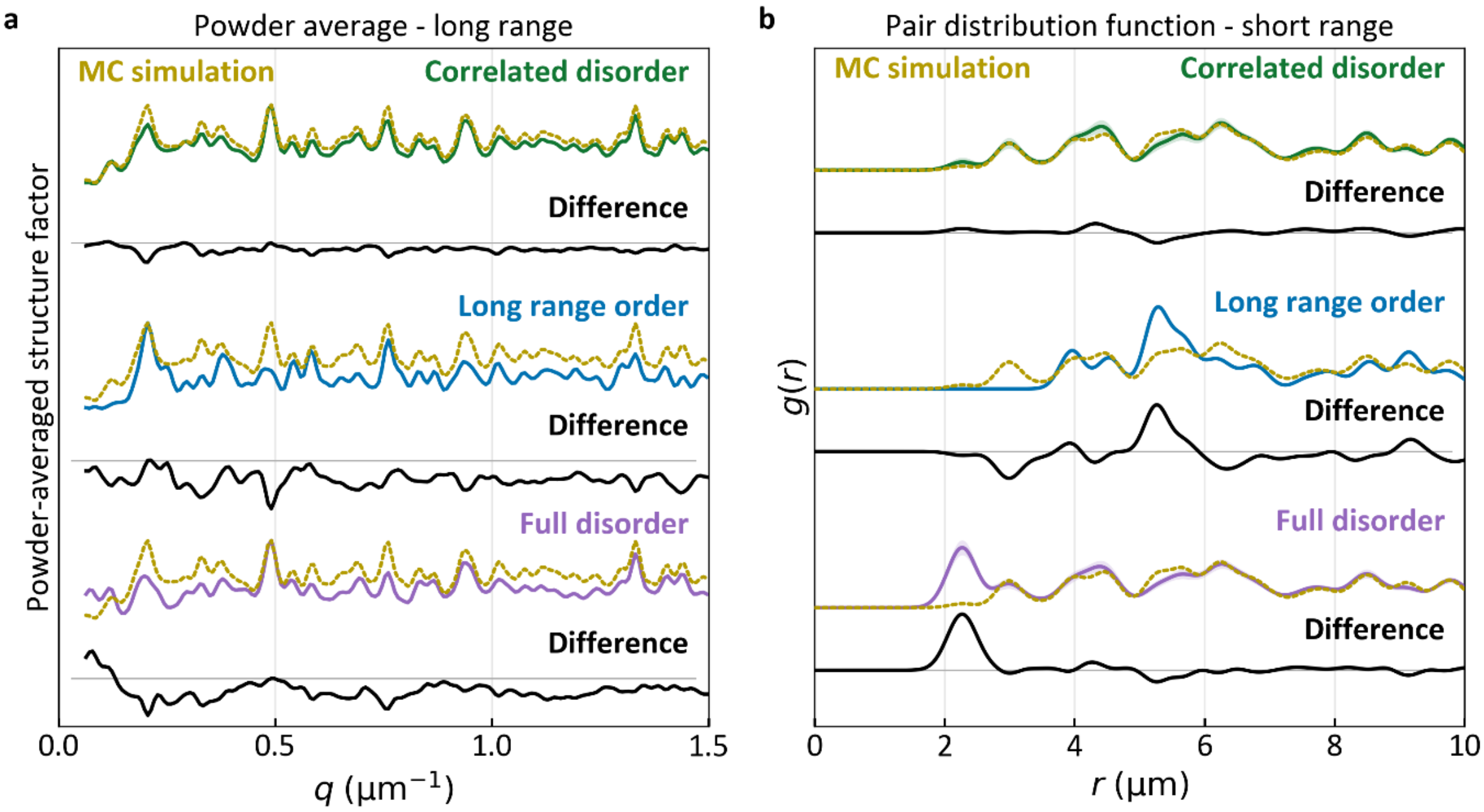

**Fig. S7 | Comparison of powder-averaged structure factors and pair distribution functions for spin states obtained with Monte Carlo simulations and the three model systems. a**, Powder-averaged structure factors, which probe long-range correlations. **b**, Pair distribution functions, which probe short-range correlations. Data from the Monte Carlo simulations are shown with a yellow dashed curve. Blue, green, and purple curves correspond to the long-range order, correlated disorder, and full disorder models, respectively. For each of these three curves, the difference in amplitude between the model curve and Monte Carlo (MC) simulation is plotted in black. All curves are on the same vertical scale.

**Estimate of the ground-state entropy**

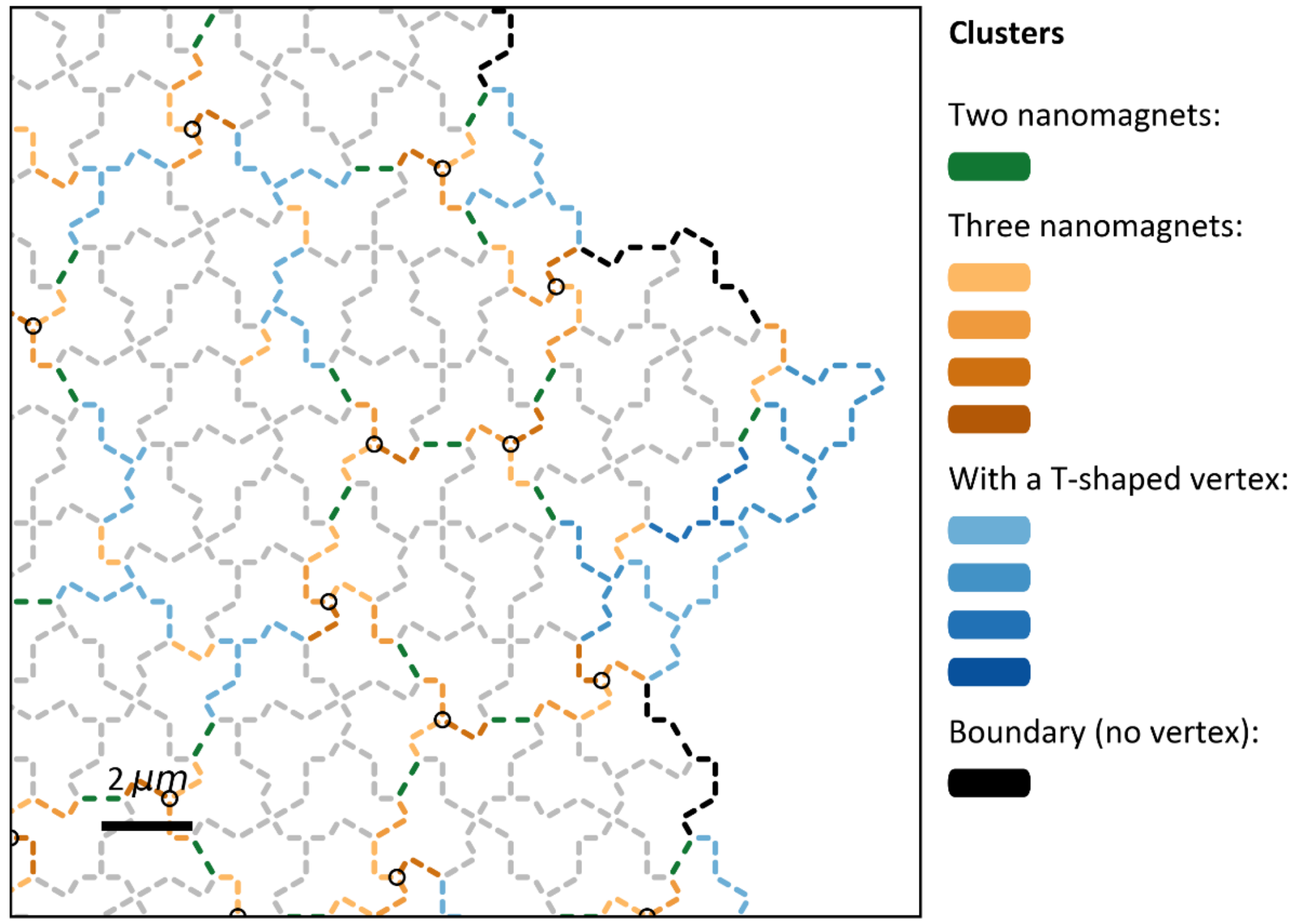

**Fig. S8 | Schematic highlighting bordering clusters.** Antihat clusters are shown in grey. Among the bordering nanomagnets, clusters of spins are defined by grouping together the nanomagnets on all the edges between Y-shaped vertices. Green nanomagnets correspond to two-nanomagnet clusters, orange nanomagnets correspond to three-nanomagnet clusters, blue nanomagnets correspond to clusters with one T-shaped vertex in the middle, and black nanomagnets correspond to boundary clusters, which contain no vertices and lie along the edge of the array of nanomagnets. Three-nanomagnet and T-vertex clusters are shown in different shades of the same colour so that adjacent clusters of the same type can be distinguished. Open circles mark the junctions where three-nanomagnet clusters meet at a Y-shaped vertex.

In the ground state of the Einstein ASI, each antihat cluster has three inequivalent ground-state configurations corresponding to the two possible positions of the excited T-shaped vertex (see Fig. S6a), which doubles to six under time-reversal symmetry.

The total number of antihat clusters is $N_{ah}$, and to each of them we therefore assign an entropy of $\ln(6)$. We then group the remaining spins between the antihats into $N_{bd}$ border clusters. As Y-shaped vertices are the only vertices with degeneracy in their ground state, the border clusters are defined as a set of edges between Y-shaped vertices. We show the nanomagnets associated with border clusters in Fig. S8, with the colours corresponding to their length or, equivalently, the number of nanomagnets. Among the three-nanomagnet border clusters (shown in different shades of orange), some connect to a Y-shaped vertex that is entirely within the boundary, indicated by white circles. There are $N_Y$ such vertices, so $3 \times N_Y$ clusters involved, as each Y-shaped vertex, highlighted by white circles, connects three border clusters. As we know that all Y-shaped vertices are in their ground state, we assign to each of these the same entropy that they have in the kagome ASI in the Kagome Ice I phase of $\ln(3/\sqrt{2})$ per vertex[1]. The rest of the border clusters are modelled as independent Ising variables to which we assign an entropy of $\ln(2)$. The total ground-state entropy is therefore given by:

$$S_0^{indep} = k_B \ln\left(6^{N_{ah}} \cdot \frac{3}{\sqrt{2}}^{N_Y} \cdot 2^{N_{bd}-3N_Y}\right).$$

For Supertile 4 with $N = 7633$ nanomagnets, we find $N_{ah} = 81$, $N_Y = 103$, and $N_{bd} = 785$ as detailed in Fig. S8: 173 two-nanomagnet clusters in green, 455 three-magnet clusters in shades of orange, 116 clusters with a T-shaped vertex in shades of blue, and 41 vertices at the boundary of the system in black. This gives

$$S_0^{indep}/N \approx 0.072\ k_B \text{ per spin}.$$

In our approximation, we suppress the correlations between border clusters beyond having Y-shaped vertices in their ground state, as well as between antihat clusters. We also neglect the fact that some Y-shaped vertices can also be found between other types of border clusters near the boundary of the system. Indeed, their contribution should vanish as the system size increases. Despite this, our estimate is very close to the numerical value (see Fig. S9 below).

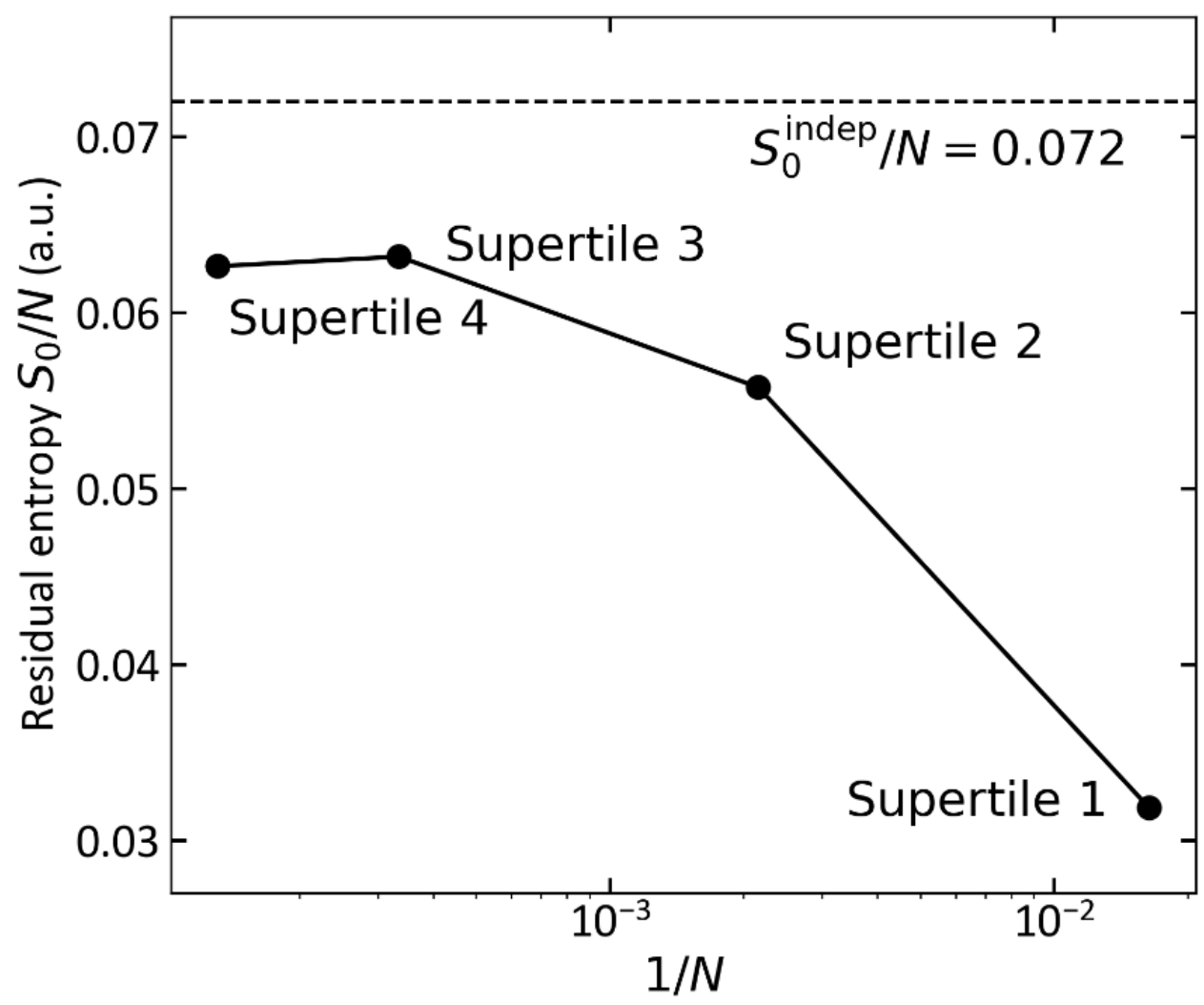

**Fig. S9 | Residual entropy per spin as a function of system size.** N is the number of macrospins in the corresponding supertile. Data points are the values obtained from Monte Carlo simulations. The estimate of the ground-state entropy associated with independent spin clusters for Supertile 4 of 0.072 is indicated by the horizontal dashed line.